\documentclass[%
 reprint,
 amsmath,amssymb,
 aps,
 prl,
]{revtex4-1}
\usepackage{graphicx}% Include figure files
\usepackage{dcolumn}% Align table columns on decimal point
\usepackage{bm}% bold math
\begin{document}
\preprint{APS/123-QED}
\title{Laser-Driven Electron Lensing in Silicon Microstructures}
\author{Dylan S. Black,$^{1*}$ Kenneth J. Leedle,$^1$ Yu Miao,$^1$ Uwe Niedermayer,$^3$ Robert L. Byer,$^2$ Olav Solgaard$^1$ }
\affiliation{%
 $^1$Stanford University, Department of Electrical Engineering,
 $^2$Stanford University, Department of Applied Physics,
 $^3$Technische Universit\"at Darmstadt, Institut f\"ur Theorie Elektromagnetischer Felder
}%
\collaboration{ACHIP Collaboration}
\date{\today}
\begin{abstract}
We demonstrate a laser-driven, tunable electron lens fabricated in monolithic silicon. The lens consists of an array of silicon pillars pumped symmetrically by two 300 fs, 1.95 $\mu$m wavelength, nJ-class laser pulses from an optical parametric amplifier. The optical near-field of the pillar structure focuses electrons in the plane perpendicular to the pillar axes. With 100 $\pm$ 10 MV/m incident laser fields, the lens focal length is measured to be 50 $\pm$ 4 $\mu$m, which corresponds to an equivalent quadrupole focusing gradient $B'$ of 1.4 $\pm$ 0.1 MT/m. By varying the incident laser field strength, the lens can be tuned from a 21 $\pm$ 2 $\mu$m focal length ($B'>3.3$ MT/m) to focal lengths on the cm-scale. 
\end{abstract}
\maketitle

The Dielectric Laser Accelerator (DLA) is a dielectric microstructure which harnesses the large electric fields in femtosecond-pulsed lasers to produce an electron linear accelerator with acceleration gradients orders of magnitude higher  than conventional metal accelerators \cite{England:14,Wootton:17}. The microstructure is a sub-wavelength grating whose optical near-field is phasematched to a propagating electron beam, thereby accelerating the electron beam. The accelerator size is commensurate with its drive wavelength; while advantageous in some respects, this presents new challenges. To confine an electron beam to a $\mu$m-scale DLA channel, a lens with focusing strength many orders of magnitude higher than currently available is necessary.

In conventional accelerators, the magnetic quadrupole is the preferred lens for charged particle focusing due to its high focusing strength, low dispersion, and linear field gradient \cite{Smith:70}. Focusing strength is defined as $k = 1/(fL)$, where $f$ is the focal length and $L$ the length of the lens. The magnetic quadrupole focusing strength is $k [\textrm{m}^{-2}] \approx {0.3 B' [\textrm{T}/\textrm{m}]}/{p\textrm{[GeV}/c]}$ \cite{Rossbach:93}. It is common to compare lens strengths by their equivalent quadrupole field gradient $B'$, and this convention is adopted throughout this letter. The required $B'$ for DLA is between 100-1000 kT/m, set by the resonant defocusing forces of the synchronous accelerating mode \cite{England:14,Ody:17,Niedermayer:18}. Conventional quadrupoles can achieve a $B'$ of only 500 T/m \cite{Strait:01,Eichner:07,Becker:09}. The other commonly employed static-field lenses, einzel lenses \cite{Read:01} and solenoids \cite{Kumar:08}, are also far too weak to achieve effective confinement. To realize an electron linac on-chip, a new type of lens, as proposed in \cite{Niedermayer:18}, must be designed.

\begin{figure}[h!]
\includegraphics[width=8.5cm]{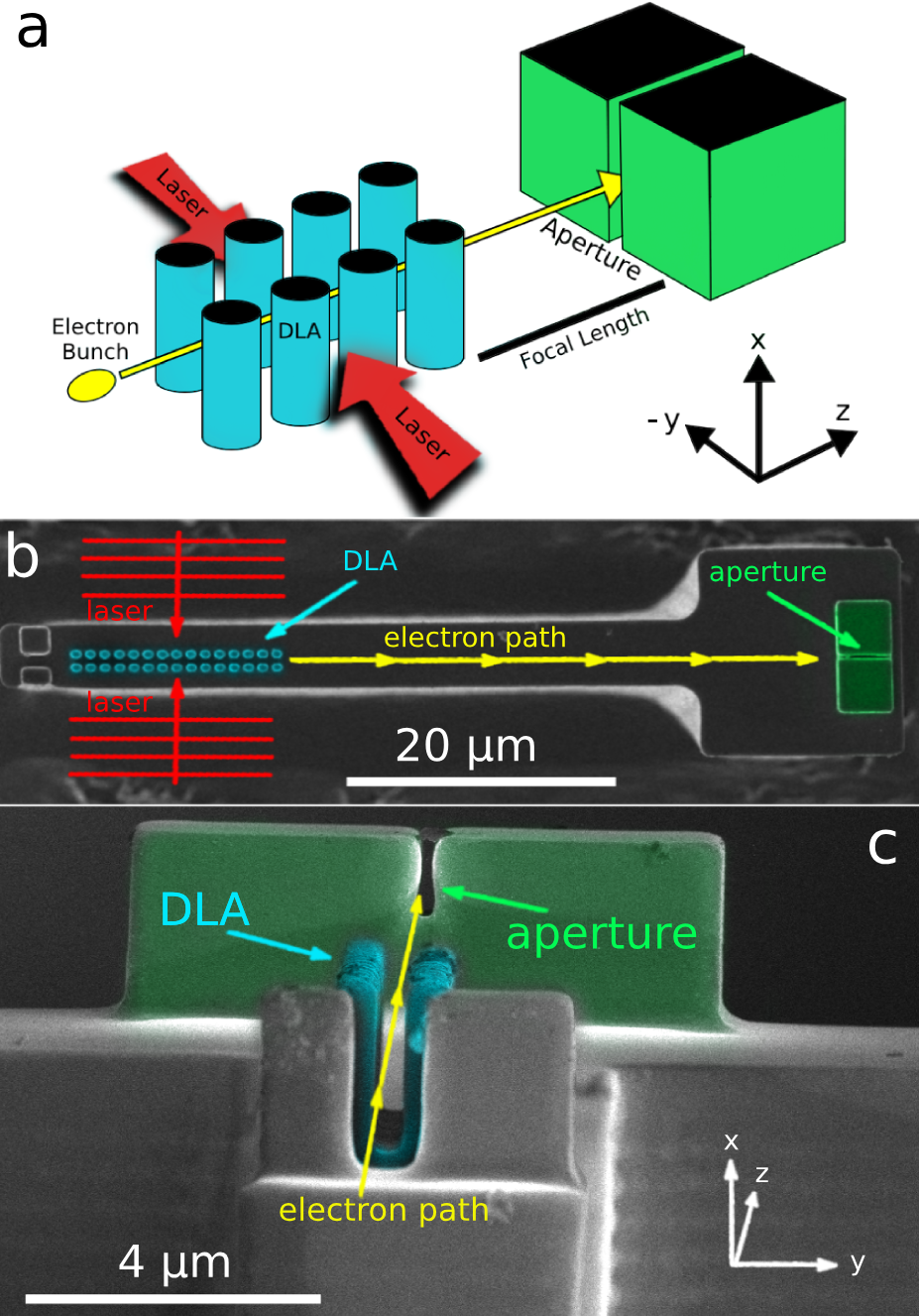}
\caption{a) An electron beam passes through a DLA lens with two identical laser pulses normally incident upon it. The beam is focused, travels approximately one focal length, and is filtered by an aperture of two silicon blocks with a small gap between them.  b) An SEM of the lens and aperture, viewed from above. The lens is composed of two rows of 15 pillars each. The drift length is 39.6 $\mu$m. c) An SEM showing the aperture structure.}
\label{fig:SEM}
\end{figure} 

The ideal lens for DLA beam confinement would be stable, high-power, tunable, and monolithically integrable into the current architecture. Monolithic integration is especially critical, as alignment tolerances for $\mu$m-scale beamlines are measured in nanometers, and such tight alignment tolerances are only realistically accessible by use of a monolithic fabrication procedure for both lens and accelerator.

Electrodynamic lenses can provide the required focusing strength. Active plasma lenses have focusing gradients exceeding 3 kT/m \cite{Tilborg:15}, while plasma wakefield lensing has focusing strengths on the order of 1 MT/m \cite{Ng:01,Chiadroni:18}. Plasma wakefield lensing fits naturally with plasma-based accelerators \cite{Kuschel:16}, and provides more than sufficient focusing strength. However, integration of plasma lenses with photonic accelerators would require generation of stable plasmas on-chip that are compatible with the accelerator nanofabrication processes.

Strong lensing effects can also be derived from the optical near-fields of femtosecond-pulsed lasers. Recently, a laser-driven lens with a 190 $\mu$m focal length was demonstrated by McNeur et al. \cite{McNeur:18}, which is estimated to have an equivalent $B'$ of 85 kT/m. The evanescent fields near a curved silicon grating generate a focusing field in the plane of the wafer. However, in single gratings there are always undesirable out-of-plane deflection forces \cite{Leedle:15}. Moreover, the curvature of the grating causes undesirable coupling of the two transverse planes, which complicates the lens implementation in a multi-stage accelerator design. 

In this letter, we demonstrate a laser-driven, solid-state electron lens based on the DLA architecture, which was first proposed by Plettner et al. in \cite{Plettner:09,Plettner:11}, and whose specific architecture is discussed by Leedle et al. in \cite{Leedle:18}. When illuminated by two laser pulses, each with an electric field of 100 $\pm$ 10 MV/m, its focal length is measured to be 50 $\pm$ 4 $\mu$m ($B' = 1.4 \pm 0.1$ MT/m). The lens strength is continuously tunable, and we demonstrate its tuning to a focal length below 21 $\pm$ 2 $\mu$m ($B' = 3.3$ $\pm$ $0.3$ MT/m). The lensing behavior agrees well with simulation, and we provide a linearized model that approximates the lens focal length. This lens architecture adds no additional complexity to the accelerator fabrication process, as it uses identical procedures and can be integrated directly into the lithographic mask. The demonstrated focusing strength is sufficient to confine an electron beam to a $\mu$m-scale beamline. We propose that this lens be used in an Alternating Phase Focusing (APF) scheme \cite{Niedermayer:18}, which allows stable beam confinement and acceleration over arbitrary distances.

The lens structure (Fig.~\ref{fig:SEM}) is fabricated from monolithic 5$-$10 $\Omega$-cm B:Si, and consists of 2 rows of 15 pillars, with periodicity $\Lambda = 1013$ nm and a 375 nm wide channel between the rows. The pillars are elliptical (613 nm x 459 nm), with a height of 2.7 $\mu$m. The electron beam passes though the central channel, where it interacts with two 300 $\pm$ 25 fs laser pulses with a center wavelength of 1.950 $\pm$ 0.005 $\mu$m and a 1/$e^2$ radius of 20 $\pm$ 2 $\mu$m. Following the lens is 39.6 $\mu$m of drift space, then an aperture consisting of two 4 $\mu$m x 4 $\mu$m x 2.7 $\mu$m silicon blocks with a gap of 150 $\pm$ 10 nm between them.

The electromagnetic fields in the lens are described following \cite{Leedle:18,NiedermayerPRAB:17,Wei:17}. We consider a dual-pillar structure semi-infinite in $x$, symmetric in $y$, and periodic in $z$. The device is illuminated by two counter-propagating $z$-polarized plane waves, incident from the $\pm y$ directions (Fig.~\ref{fig:SEM}), each with electric field $E_\textrm{inc}$.  The electrons travel in $z$ with velocity $\beta = v/c$. The synchronicity (or phasematching) condition between the laser field and the electron is
\begin{equation}
    \beta \lambda_0 = \Lambda,
    \label{eq:phasematching}
\end{equation} where $\lambda_0$ is the central laser wavelength and $\Lambda$ is the structure periodicity. The Lorentz force on an electron inside the structure, assuming Eq.~\ref{eq:phasematching} is satisfied and non-phasematched harmonics are negligible, is
\begin{widetext}
\begin{equation}
\vec{F} = -\frac{q e_1}{2 \gamma} \textrm{Re}
\begin{bmatrix} 
0 \\
\sin\phi \left[(e^{i \theta} - 1) \cosh(k_y y)+(e^{i \theta} + 1) \sinh(k_y y)\right]\\
\gamma \cos\phi \left[(e^{i \theta} + 1) \cosh(k_y y)+(e^{i \theta} - 1) \sinh(k_y y)\right]\\
\end{bmatrix}
\label{eq:totalForceEquation}
\end{equation}
\end{widetext}
where $\gamma = (1-\beta^2)^{-1/2}$, $k_y = 2 \pi/\beta \gamma \lambda_0$ is the wavevector of the evanescent field, $q$ is the elementary charge, and $e_1$ is the magnitude of the synchronous accelerating field at $y=0$ \cite{NiedermayerPRAB:17}. We assume a laser phase such that $e_1$ is purely real and positive, and define the structure constant $c_s = e_1 / E_\text{inc}$. $\phi$ is the phase of the electron relative to the optical cycle of the $+y$ plane wave, and $\theta$ is the relative phase between the counter-propagating waves. The force along the $x$ coordinate is zero by the semi-infinite assumption. Previous experimental results, as well as 3-D FDTD simulations, indicate that the semi-infinite approximation works well for 2.7 $\mu$m tall (or taller) pillars \cite{Leedle:18}. The magnitude of the transverse and longitudinal forces differ by a factor of $\gamma$, as expected from the Panofsky-Wenzel theorem \cite{Panofsky:56}. 

For in-phase drive lasers ($\theta$ = 0) and assuming perfect synchronicity, the focal length of a device with $N$ periods is approximately
\begin{equation}
f \approx \frac{\beta^2 \gamma^3 m_e c^2}{2 \pi N q e_1 \sin\phi}
\label{eq:focLenLin}
\end{equation}
Eq.~\ref{eq:focLenLin} is valid for a sufficiently small $N$ such that the thin-lens approximation holds. The lens strengths considered here restrict the validity of Eq.~\ref{eq:focLenLin} to devices with $N<18$. 

Neglecting phase slippage due to acceleration, valid for short structures, the energy gain in the $\theta$ = 0 mode is \begin{equation}
    \Delta U \approx -q e_1 N \Lambda \left(\frac{\cos\theta+1}{2}\right) \cos\phi
    \label{eq:energyGain}
\end{equation}

For out-of-phase drive lasers ($\theta = \pm\pi$), there exists, to first order in $y$, a constant deflection force whose direction varies sinusoidally with $\phi$. Further discussion of the accelerator modes is contained in the Supplementary Material.

The electron bunch is modeled as a collection of normally distributed $x$, $y$, and $\phi$ values. Each electron experiences a focal length drawn from the distribution of Eq.~\ref{eq:focLenLin}, and for an electron beam much longer than an optical cycle ($\sim$6 fs), the electrons within the bunch stochastically sample all possible focal lengths. To measure the minimum lens focal length, a very small aperture is placed one focal length from the lens and acts as a temporal filter, biasing electron detection towards the focusing phases ($0< \phi < \pi $) over the defocusing phases ($\pi< \phi < 2\pi$).

The electron beam used in this experiment was produced with a 300 $\pm$ 25 fs FWHM, 100 kHz, 256 nm laser pulse incident on a flat copper cathode. The electron beam has a circular gaussian spatial profile and a 4$\sigma$ width of 780 $\pm$ 63 nm at the lens entrance, measured by a knife edge scan. The geometric 1-D emittance is estimated to be $\sim$0.5 nm rad. The beam energy is 89.4 $\pm$ 0.1 keV, which corresponds to $\beta \approx 0.525$. The beam current is set to 730 $\pm$ 200 $e^-$/s to avoid energy broadening from space charge effects at the cathode. Each electron pulse at the interaction point is 740 $\pm$ 110 fs FWHM in length, measured by cross correlation with a 300 $\pm$ 25 fs laser pulse.  

\begin{figure}[]
\includegraphics[width=8.5cm]{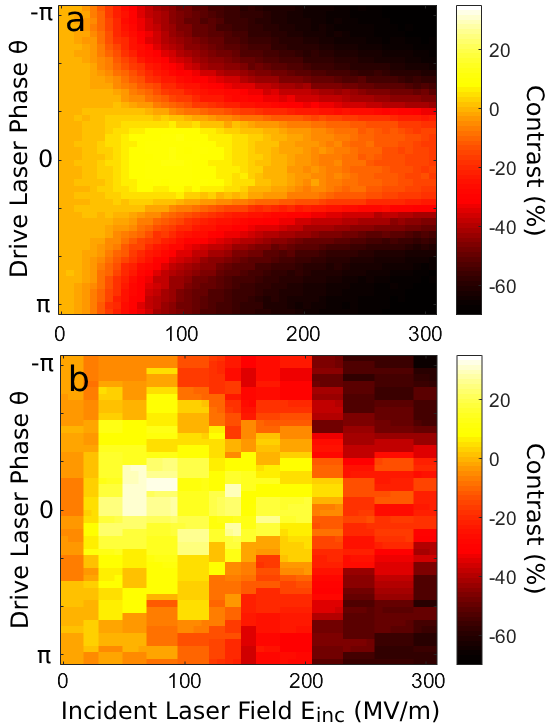}
\caption{a) Simulated contrast plotted against relative drive laser phase $\theta$ and electric field $E_\text{inc}$. The simulation is a symplectic 2-D particle-tracking scheme based on \cite{NiedermayerPRAB:17} which applies a momentum kick once per lens period equal to the time integral of Eq.~\ref{eq:totalForceEquation} over one structure period. Expected contrast is calculated by a Monte Carlo approach. b) Contrast is measured as a function of $\theta$ and  $E_\text{inc}$.}
\label{fig:SimulationExperiment}
\end{figure}

43 $\pm$ 8\% of electrons are transmitted through the aperture with the laser (and thus the lens) off. Leakage through the silicon blocks is small; the blocks block 95\% of incident electrons. For the $\theta = 0$ focusing mode, an increase in electron transmission through the aperture is expected, with maximal transmission when the drift length is matched to the lens focal length. ``Contrast," the percent increase in electron transmission when the lens is turned on (Eq.~\ref{eq:contrast}), quantifies this increase.
\begin{equation}
\textrm{Contrast} [\%] \equiv 100\left(\frac{T_\text{on}}{T_\text{off}}-1\right)
\label{eq:contrast}
\end{equation}
$T_\textrm{on}$ is the electron transmission with the lens on, and $T_\textrm{off}$ is the electron transmission with the lens off. After the aperture, the electrons travel through a magnetic spectrometer with an energy resolution of 100 eV, and are detected on a microchannel plate detector.

Electron transmission simulations were carried out for a range of incident laser fields ($E_\text{inc}$) and drive laser phases ($\theta$) (Fig.~\ref{fig:SimulationExperiment}a). Due to the long bunch length, the increased transmission from the focusing phases is partially offset by the decreased transmission from the defocusing phases. Thus, the expected contrast in this operating mode is low, only 11\%. However, for the $\theta = \pm \pi$ mode, a large transmission decrease for all values of $\phi$ is expected. 

\begin{figure}[]
\includegraphics[width=8.5cm]{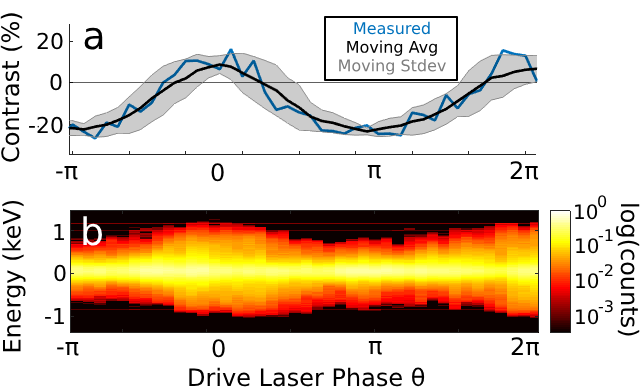}
\caption{a) The measured contrast as a function of $\theta$ for $E_\textrm{inc}$ = 137 $\pm$ 13 MV/m. The blue line is measured data, the black line and gray shaded area are the moving average and standard deviation, respectively. b) The electron energy spectrum is measured simultaneously with the phase sweep in a). Electron counts are normalized to the maximum value.}
\label{fig:PhaseSweep}
\end{figure} 

In Fig.~\ref{fig:SimulationExperiment}b, the parameter space simulated in Fig.~\ref{fig:SimulationExperiment}a is measured. The experimental data agrees qualitatively with the simulation. There is a small contrast peak at $E_\text{inc} = $ 100 $\pm$ 10 MV/m for $\theta = 0$, with a large region of strong negative contrast in the $\theta = \pm \pi$ region.

The energy modulation and contrast as a function of drive laser phase is shown in Fig.~\ref{fig:PhaseSweep}. The sinusoidal variation of energy gain and contrast predicted by Eq.~\ref{eq:energyGain} is demonstrated, and the peak transmission and peak energy modulation occur at the same $\theta$, in agreement with theory. 

\begin{figure}[]
\includegraphics[width=8.5cm]{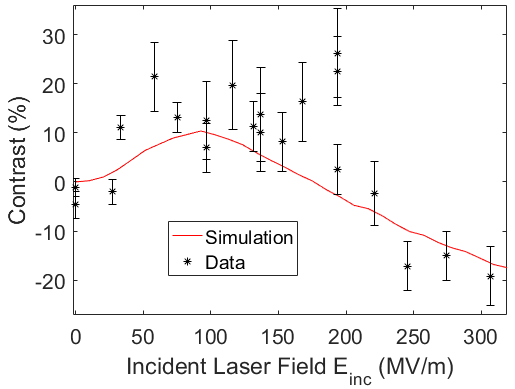}
\caption{This figure plots the contrast and standard deviation as a function of $E_\textrm{inc}$ for $\theta = 0$. The simulation curve is a cross section of Fig.~\ref{fig:SimulationExperiment}a at $\theta = 0$. Duplicate runs have been included.} 
\label{fig:ExperimentLine}
\end{figure}

Fig.~\ref{fig:ExperimentLine} plots contrast in the $\theta = 0$ focusing mode as a function of $E_\text{inc}$, to aid a visual comparison between simulation and experiment. The co-location of the peak contrast for simulation and experiment is apparent. Duplicate runs omitted from Fig.~\ref{fig:SimulationExperiment}b for visual clarity are included in Fig.~\ref{fig:ExperimentLine}. 

The contrast peak at $E_\text{inc} =$ 100 $\pm$ 10 MV/m ($e_1 =$ 38  MV/m) corresponds to a focal length of 64 $\pm$ 6 $\mu$m in the linearized approximation (Eq.~\ref{eq:focLenLin}). Our experimentally measured focal length, defined as the total distance from the lens principal plane to aperture center, is measured to be 50 $\pm$ 4 $\mu$m ($B' = 1.4 \pm 0.1$ MT/m). The measured focal power is greater than predicted by Eq.~\ref{eq:focLenLin}, indicating that the thin-lens approximation breaks down at these lens strengths. The simulation, which uses the forces from Eq.~\ref{eq:totalForceEquation}, accurately predicts the incident field which gives peak contrast. Because the measured focal length will always be less than that predicted by the linearized approximation, Eq.~\ref{eq:focLenLin} can be considered a lower bound on the lens focusing strength. The incident laser field is increased to a maximum of 306 $\pm$ 16 MV/m, corresponding to a linearized focal length of 21 $\pm$ 2 $\mu$m ($B' = 3.3 \pm 0.3$ MT/m). 

The structure constant $c_s$ was measured to be 0.38 $\pm$ 0.04 for this structure, with a maximum acceleration gradient $e_1$ of 111 $\pm$ 6 MeV/m. Previous work with similar structures has demonstrated $c_s = 0.27 \pm 0.03$, with $e_1 = 133 \pm 8$ MeV/m \cite{Leedle:18}.

The main experimental limitation was electron beam pointing instability. The electron beam could be stably aligned to the aperture for approximately 60 s, limited by thermal drifts. An averaging time of 3 s per data point ($\sim$2000 electrons) was chosen to compromise between high frequency and low frequency noise. The effects of random beam motion are partially compensated by normalizing the total number of transmitted electrons to the fraction of electrons detected at the initial beam energy of 89.4 keV. Since the laser pulse is shorter than the electron pulse, many electrons within the pulse do not interact with the laser, and these electrons are all detected at 89.4 keV. The fluctuation of the electron counts detected at 89.4 keV serves as an instantaneous measure of electron beam misalignment. The details are included in the Supplementary Material.

The focusing strength scales as $1/\gamma^3$, which is a less favorable scaling than the $1/\gamma^2$ scaling of solenoids or the $1/\gamma$ scaling of quadrupoles. The equivalence point is found by equating $k$ of a quadrupole lens to $k$ of a DLA lens and solving for $\gamma$. The equivalence point here is $\sim$35 MeV. This can be increased by increasing either the electric field or the lens length. Efficient laser-electron interaction requires that the lens material refractive index $n$ be greater than $1/\beta$ \cite{Hughes:18,Kozak:17}. Silicon has $n \approx 3.45$ at $\lambda_0 = $ 2.0 $\mu$m, corresponding to a lower limit of $\beta \approx 0.29$ (23 keV).

The lens focal length is continuously variable from approximately 20 microns to the cm-scale. However, electron pulses of duration $\tau_p << \lambda_0/c$ are required for use as a single focal length lens. Fortunately, a pulse train created using the same DLA architecture \cite{Niedermayer:18,NiedermayerJournPhysConf:17} has the correct microbunch length and periodicity. The use of evanescent fields to focus electrons necessitates a narrow aperture, and so its emittance acceptance is small. The beam emittance in this lens is not conserved, however the emittance growth due to field nonlinearity is quite small. Achievable spot sizes are limited by third-order aberrations from the sinh focusing profile. Lens nonlinearities are treated more fully by Niedermayer et al. in \cite{NiedermayerPRAB:17,Niedermayer:18}. 

We propose to use this lens in an Alternating Phase Focusing (APF) confinement scheme. Briefly, lens stages are alternated with drift sections chosen to provide a specific phase offset between lens stages. For example, a drift length of one half-period is equivalent to a $\pi$ phase delay in $\phi$, which reverses the sign of the lens focal length. If the phase offsets are chosen appropriately, it is possible to achieve stable confinement in both the transverse and longitudinal directions simultaneously, which can then be combined with high gradient acceleration, as detailed in \cite{Niedermayer:18}. The confinement requirements for DLA are set by the resonant defocusing forces \cite{England:14,Ody:17,Niedermayer:18}, and since the defocusing forces are exactly those forces described by Eq.~\ref{eq:totalForceEquation}, the lensing forces presented here have precisely the same strength as the resonant defocusing itself. Thus, using this architecture, the focusing strength requirement for confinement in DLA is satisfied automatically, even for the large defocusing forces present in high-gradient accelerators \cite{Cesar:18}. 

In summary, we have demonstrated a laser-driven, continuously tunable electrodynamic lens with a focusing strength equivalent to those observed in plasmas \cite{Ng:01,Chiadroni:18} and which far exceeds any static-field lens. Its design is easily and monolithically integrable with current photonic accelerator architectures, and its strength is sufficient to confine an electron beam to an accelerator channel less than 1 $\mu$m wide for an arbitrary distance \cite{Niedermayer:18}. This removes a major roadblock in the development of scalable on-chip electron accelerators. 

%\bibliography{FullBibliography}

\bibliographystyle{unsrtnat} 

%%%%%%%%%% Merge with supplemental materials %%%%%%%%%%
\pagebreak

\begin{center}
\textbf{\large Supplementary Material}
\end{center}
%%%%%%%%%% Merge with supplemental materials %%%%%%%%%%
%%%%%%%%%% Prefix a "S" to all equations, figures, tables and reset the counter %%%%%%%%%%
\setcounter{equation}{0}
\setcounter{figure}{0}
\setcounter{table}{0}
\setcounter{page}{1}
\makeatletter
\renewcommand{\theequation}{S\arabic{equation}}
\renewcommand{\thefigure}{S\arabic{figure}}
\renewcommand{\bibnumfmt}[1]{[S#1]}
\renewcommand{\citenumfont}[1]{S#1}
%%%%%%%%%% Prefix a "S" to all equations, figures, tables and reset the counter %%%%%%%%%%

\section{Summary}
Supplementary material for \textit{Laser-Driven Electron Lensing in Silicon Microstructures}, by Black et al. This appendix contains derivations of some equations stated in the main paper, as well as the details of the normalization scheme used to process the reported experimental data. All references refer to the bibliography in the main paper. 

\section{DLA Lens Focal Length}

For convenience, the definitions of various parameters are restated here.

As in the main manuscript, we consider a dual-pillar structure semi-infinite in $x$, symmetric in $y$, and periodic in $z$ (following [20-22]). The device is illuminated by two counter-propagating $z$-polarized plane waves, incident from the $y$ direction.  The electrons travel along $z$ with velocity $\beta = v/c$, $\lambda_0$ is the laser center wavelength, $\Lambda$ is the structure periodicity, $\gamma = (1-\beta^2)^{-1/2}$, $k_y = 2 \pi/\beta \gamma \lambda_0$ is the wavevector of the evanescent field, $q$ is the elementary charge, $e_1$ is the magnitude of the synchronous accelerating field at $y = 0$, $\phi$ is the phase of the electron relative to the laser optical cycle of the $+y$ plane wave within the periodic structure, and $\theta$ is the relative phase between the counter-propagating waves. We assume a laser phase such that $e_1$ is purely real and positive. 

Beginning from the Lorentz force (Eq. 2 in the main paper), the transverse force $F_y$ seen by an electron at a given $\phi$ can be written as a superposition of a cosh term and a sinh term, which correspond to the deflecting and focusing modes, respectively. \begin{equation}
    F_y \propto \textrm{Re}\left[\frac{(e^{i \theta} - 1)}{2} \cosh(k_y y)+\frac{(e^{i \theta} + 1)}{2} \sinh(k_y y)\right]
\end{equation}
The relative drive laser phase $\theta$ controls the behavior of the device. For in-phase drive lasers ($\theta$ = 0) and assuming perfect synchronicity, integrating $F_y$ in time across one period $\Lambda$ (see [21]) gives
\begin{equation}
	\Delta p_y = -\frac{\Lambda}{\beta c}\frac{q e_1}{\gamma} \sinh\left(\frac{2 \pi}{\beta \gamma \lambda_0} y\right) \sin\phi 
\end{equation} 
Applying $\beta \lambda_0 = \Lambda$ and expanding to first order in $y$ yields
\begin{equation}
	\Delta p_y \approx -\frac{2 \pi q e_1}{c \beta \gamma^2} \frac{}{} y   \sin\phi
\end{equation} 
The angular deviation of the electron trajectory over one DLA period $\Delta y'_1$ near the center of the channel is 
\begin{equation}
\Delta y'_{1} = \frac{\Delta p_y}{p_{z0}} \approx -\frac{2 \pi q e_1 \sin\phi}{\beta^2 \gamma^3 m_e c^2}y
\label{eq:AngKickLin}
\end{equation}
For a sufficiently small number of periods $N$ such that $y$ is nearly constant, the total angular deflection is $\Delta y' \approx N \Delta y'_{1}$. In analogy with the first order approximation of lens focal length, $\Delta y' = -y/f$, the focal length of the DLA lens is approximately, 
\begin{equation}
f \approx \frac{\beta^2 \gamma^3 m_e c^2}{2 \pi N q e_1 \sin\phi}
\label{eq:focLenLin}
\end{equation}

Eq.~\ref{eq:focLenLin} is valid only for cases where the focal length $f$ is much less than the length of the lens. For a lens of $N$ periods, the lens length is $N\Lambda$ and the thin lens approximation holds when \begin{equation}
    N^2 \leq \frac{\beta^2 \gamma^3 m_e c^2}{2 \pi \Lambda q e_1}
\end{equation}

For the lens strengths considered here, the lens can be considered ``thin" if $N < 18$. Longer lenses must be modeled as thick lenses (see [6]).

\section{Energy Gain of Short DLA Structures}

On-axis ($y = 0$), the electron sees a longitudinal force $F_z$ of \begin{equation}
    F_z = \textrm{Re} \left[ -\frac{q e_1}{2} (e^{i \theta} + 1) \cos\phi \right]
\end{equation}
Since energy gain is small for small $N$, phase slippage due to acceleration can be neglected, and the total energy gain over $N$ periods is found by integrating $F_z$ over the length of the structure, yielding an on-axis energy gain of \begin{equation}
    \Delta U \approx -q e_1 N \Lambda \left(\frac{\cos\theta+1}{2}\right) \cos\phi
\end{equation}

\section{Deflection Forces}

For $\theta = \pi$, the forces in the transverse direction are given by \begin{equation}
    F_y = \frac{q e_1}{\gamma} \cosh(k_y y) \sin \phi 
\end{equation} 
In the limit of small deviations from the $y=0$ axis, $F_y$ is \begin{equation}
    F_y \approx \frac{q e_1}{\gamma} \left(1 + \frac{1}{2}\left(\frac{2 \pi y}{\beta \gamma \lambda_0}\right)^2\right) \sin \phi 
\end{equation} 
For a lens with a channel width much less than $\lambda_0$, higher order terms may be neglected and the deflection force is approximately constant in magnitude across the lens channel. The direction of the deflection force varies sinusoidally with $\phi$, and the magnitude is directly proportional to the laser field strength via $e_1$. Operated in this mode, the structure can be used as an optical frequency streaking element.

The angular deflection for a single period is easily calculated by integration of $F_y$ over one period, yielding, to leading order,

\begin{equation}
\Delta y'_{1} \approx \frac{q e_1 \Lambda}{\beta^2 \gamma^2 m_e c^2} \sin \phi
\label{eq:AngKickSinh}
\end{equation}

Since the deflection force is constant to leading order in $y$, angular deflection for $N$ periods is well approximated by \begin{equation}
    \Delta y' \approx \frac{q e_1 N \Lambda}{\beta^2 \gamma^2 m_e c^2} \sin \phi
\end{equation}

The same caveats that limit the applicability of the thin-lens approximation in the previous sections also apply here, with the additional constraint that a beam deflected uniformly by each period of accelerator structure will hit the channel wall after some distance, limiting the maximum angular deflection produced by this structure.

\section{Peak Normalization}

\subsection{Introduction}
The electron beam misalignment caused by slow thermal drifts is the main source of experimental noise. A common technique for removing external effects from a measured signal is to compare the experiment to a simultaneous reference experiment. In light optics, for example, a beamsplitter can be used to create a reference beam against which the experiment may be compared. 

An identical reference electron beam cannot be created for this experiment. However, since the laser pulse used to modulate the energy of the electron pulse is shorter than the electron pulse itself, there are many electrons detected which do not interact with the laser at all. The fluctuation of the non-interacting electron counts is not dependent on the laser fields, and can therefore be used as an instantaneous reference for beam alignment to the lens channel. 

The non-interacting electrons are all detected at the initial beam energy (see Fig. 1). Thus, by filtering the detected electrons with a magnetic spectrometer and normalizing the total electron counts to the electron counts at the initial beam energy, it is possible to partially separate the electron count fluctuations due to the laser fields from the fluctuations due to beam misalignment. In other words, the height of the central energy peak (Fig. 1) constitutes an instantaneous pseudo- reference beam against which total electron transmission can be compared. Not all of the electrons which appear at the central energy peak are non-interacting. But, since the electron pulse used in this experiment is nearly twice as long as the laser pulse, the majority of the electrons detected at the initial beam energy are, in fact, non-interacting. 

\subsection{A Note on the Experimental Procedure}

The independent variables in the experiment are the peak electric field of one drive laser $E_{\text{inc}}$, and the relative phase between the two drive lasers $\theta$. The dependent variable is the electron counts measured on the detector. Each data point taken in the experiment was taken as part of a ``phase sweep," which consists of holding $E_{\text{inc}}$ constant while $\theta$ is varied linearly, and continuously measuring the electron energy spectrum as a function of phase (see Fig. 4 in the main paper or Fig. 3 in the Supplementary Material for an example of a phase sweep). During the phase sweep, individual data points are collected, consisting of an electron energy spectrum with an averaging time of 3 s ($\sim$2000 electrons). After each phase sweep is completed, the laser is turned off, and the energy spectrum is again measured. Then, $E_{\text{inc}}$ is increased by a fixed amount, and another phase sweep is taken. In this way, the entire $(\theta, E_{\text{inc}})$ parameter space is sampled for both laser-on and laser-off conditions.

\begin{figure}[!ht]
    \begin{center}
    \includegraphics[width=\linewidth]{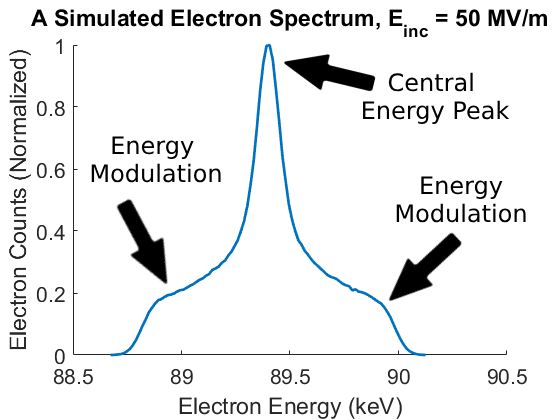}
    \caption{An example (simulated) energy spectrum which shows a large peak at the central beam energy of 89.4 keV. This spectrum was simulated using a 300 fs intensity FWHM laser pulse, with peak electric field values of 50 MV/m, and a 730 fs FWHM electron bunch. The large central peak, which is composed mainly of non-interacting electrons due to the long electron bunch, is contrasted by the smaller wings, which are composed of electrons accelerated and/or deflected by the laser fields.}
    \end{center}
    \label{figure:peak}
\end{figure}

\subsection{Notation}

Let the total number of electron counts (``transmission") measured at an energy $\epsilon$ for some $(\theta, E_{\text{inc}})$ be denoted by $T_\epsilon(\theta,E_{\text{inc}})$. In order to reduce visual clutter, this quantity will be written simply as $T_\epsilon$ when the functional dependence is not relevant. This experiment depends on measuring the ``contrast" between a laser-on condition and a laser-off condition. These two conditions will be specified by a superscript when it is necessary to distinguish them, e.g. $T^\text{on}$ (the transmission with the laser on) or $T^\text{off}$ (the transmission with the laser off), but the superscript will be omitted when the equation can be applied equally to both laser conditions, i.e. an equation involving $T$ is valid for both $T^\text{on}$ and $T^\text{off}$. The superscript will also be used to explicitly denote quantities which are measured or simulated when it is necessary to distinguish the two, e.g. $T^\text{on, sim}$ is the simulated transmission $T$ for the laser-on condition.

Normalized quantities in the simplified scheme will be written with a hat, e.g. $\hat{T}$. Corrected, normalized quantities will be written with a tilde, e.g. $\tilde{T}$. 

\subsection{Simplified Normalization Scheme}

Define the normalization factor $N(\theta,E_\text{inc})$ as \begin{equation}
     N(\theta,E_{\text{inc}}) = \frac{ T_{\epsilon_0}(\theta,E_{\text{inc}}) } { \left<T_{\epsilon_0}(\theta,E_{\text{inc}})\right>_\theta} \xrightarrow{} N = \frac{ T_{\epsilon_0} } { \left<T_{\epsilon_0}\right>_\theta}
     \label{eq:NormalizationFactor}
\end{equation} 
where $\epsilon_0$ is the central beam energy, and $\left<\right>_\theta$ is the average over all drive laser phases $\theta$ for a single phase sweep. $N$ is then a function of $\theta$ and $E_{\text{inc}}$, defined for each data point, whose mean over a single phase sweep is 1, and whose fluctuations are directly proportional to the electron counts detected with energy $\epsilon_0$. 

Define the total electron transmission $T$ as
\begin{equation}
    T =  \sum_{\epsilon}^{} T_\epsilon
\end{equation} and the normalized transmission $\hat{T}$ as \begin{equation}
    \hat{T} = T/N
\end{equation}

Contrast $C$ is defined as \begin{equation}
    C = 100\left(\frac{T^\text{on}}{T^\text{off}}-1\right)
\end{equation}

The normalized contrast $\hat{C}$ is then \begin{equation}
     \hat{C} = 100\left(\frac{\hat{T}^\text{on} }{\hat{T}^\text{off} }-1\right)
\end{equation} 

\subsection{Corrected Normalization Scheme}

The simplified normalization scheme above warps the parameter space (see Fig. 2), preventing a legitimate comparison of simulated and measured data. The normalization scheme must be corrected such that the parameter space in the zero-noise case remains invariant under normalization. 

\begin{figure}[!ht]
    \begin{center}
    \includegraphics[width=\linewidth]{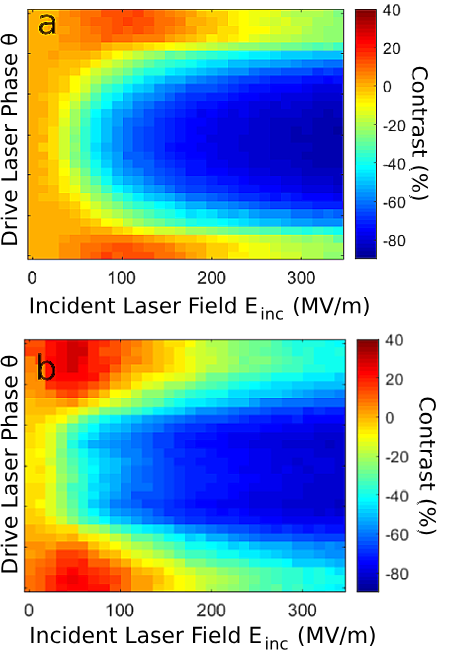}
    \caption{a) An example simulation of the experimental parameter space. b) The same parameter space with the simplified peak normalization scheme applied, showing a clear distortion of the space. Here, simulation parameters have been deliberately chosen to exaggerate the warping caused by the simplified peak normalization, but the distortion is present for any choice of parameters when the simplified normalization scheme is applied. This is then corrected to obtain the data used in the main manuscript.}
    \end{center}
    \label{warp}
\end{figure}

To construct the corrected normalized transmission $\tilde{T}$, the normalization factor $N$ is simulated for the entire parameter space, and an inverse normalization is applied to both the simulated and the measured data, i.e. for simulated data, \begin{equation}
    \tilde{T}^\text{sim} = N^\text{sim} \hat{T}^\text{sim} =  \left(\frac{N^\text{sim}}{N^\text{sim}}\right) T^\text{sim} = T^\text{sim}
\end{equation}
and for measured data, \begin{equation}
    \tilde{T}^\text{meas} = \left(\frac{N^\text{sim}}{N^\text{meas}}\right) T^\text{meas}
\end{equation}

Clearly, for simulated data, the corrected, normalized transmission $\tilde{T}$ is identical to the actual electron transmission $T$. For measured data, this is not necessarily true. In the following section, the desired signal, the ``true" transmission $T$, which is assumed to be identical to the simulated value, will be explicitly separated from any undesirable random fluctuations in the real experiment, denoted by $Z$. Define \begin{equation}
    T^\text{meas} = T + Z
\end{equation}

The measured normalization factor $N^\text{meas}$ is  \begin{equation}
    N^\text{meas} = 
    \frac{T_{\epsilon_0} + Z_{\epsilon_0}}
    {\left<T_{\epsilon_0} + Z_{\epsilon_0}\right>_\theta}
\end{equation}
The mean value of the drift over time is uncorrelated with drive laser phase $\theta$. Assuming that $Z$ averages to zero over many periods of $\theta$, it can be neglected in the denominator, giving \begin{equation}
    N^\text{meas} = 
    \frac{T_{\epsilon_0} + Z_{\epsilon_0}}
    {\left<T_{\epsilon_0}\right>_\theta}
\end{equation}

Again, to reduce visual clutter, the transmission ratios $R, \hat{R}$ will be defined as \begin{equation}
    R = \frac{T^\text{on}}{T^\text{off}}, \hat{R} = \frac{\hat{T}^\text{on}}{\hat{T}^\text{off}}
\end{equation}

The measured, normalized transmission ratio $\hat{R}^\text{meas}$ is, explicitly \begin{equation}
    \hat{R}^\text{meas} = \frac{T^\text{on, meas}/N^\text{on, meas}}{T^\text{off, meas}/N^\text{off, meas}} = \frac
     {(T^\text{on}+Z^\text{on})\left(\frac{T^\text{off}_{\epsilon_0} + Z^\text{off}_{\epsilon_0}}
    {\left<T^\text{off}_{\epsilon_0} )\right>_\theta}\right)}
     {(T^\text{off}+Z^\text{off})\left(\frac{T^\text{on}_{\epsilon_0} + Z^\text{on}_{\epsilon_0}}
    {\left<T^\text{on}_{\epsilon_0} \right>_\theta}\right)}
\end{equation}

\begin{figure}[!ht]
    \begin{center}
    \includegraphics[width=8.3cm]{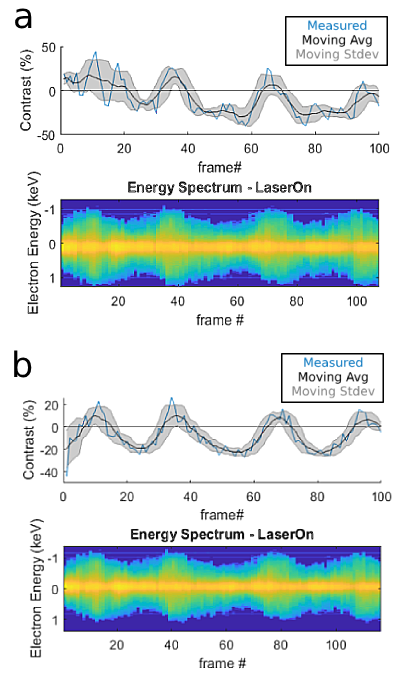}
    \caption{a) A measured phase sweep, with contrast and energy spectrum plotted as a function of frame number (3 s per frame, $\sim$2000 electrons). The slow drift of the electron beam away from the aperture manifests as the slow decay of the contrast curve. Shorter time scale fluctuations, seen in frames 0 - 20, are also present. The maximum energy modulation provides a good independent reference for the shape of the contrast curve, i.e. the contrast peaks should appear with the same frequency and at the same time as the energy modulation maxima. b) The same measured data as in panel a), with the corrected normalization scheme applied. The slow drift of the beam is corrected, yielding a sinusoidal curve with a constant average value over each period. The short time scale fluctuations are also corrected by the normalization scheme. The sinusoidal oscillation of contrast is well matched to the frequency and phase of the energy spectrum oscillation. }
    \end{center}
    \label{phaseSweep}
\end{figure}

Applying the simulated correction factor according to Eq.~\ref{eq:NormalizationFactor} gives the corrected, normalized transmission ratio $\tilde{R}^\text{meas}$ \begin{equation}
     \tilde{R}^\text{meas} = \hat{R}^\text{meas}\left(\frac{N^\text{on, sim}}{N^\text{off, sim}}\right) =
     \frac{
            (T^\text{on}+Z^\text{on})(T^\text{off}_{\epsilon_0} + Z^\text{off}_{\epsilon_0})
            \left(T^\text{on}_{\epsilon_0}\right)
            }
            {
                (T^\text{off}+Z^\text{off})
            (
                    T^\text{on}_{\epsilon_0} + Z^\text{on}_{\epsilon_0}
            )
                (T^{\text{off}}_{\epsilon_0}
            )
        } 
\end{equation}

Which can be rewritten in terms of the ``true" transmission ratio $R = T^\text{on}/T^\text{off}$ as \begin{equation}
    \tilde{R}^\text{meas} = R
     \left(\frac{1+Z^\text{on}/T^\text{on}}{1+Z^\text{on}_{\epsilon_0}/T^\text{on}_{\epsilon_0}}
     \right)\left(\frac{1+Z^\text{off}_{\epsilon_0}/T^\text{off}_{\epsilon_0}}
     {1+Z^\text{off}/T^\text{off}}\right) 
\end{equation}

where each quantity $R(\theta,E_{\text{inc}}), Z(\theta,E_{\text{inc}}), T(\theta,E_{\text{inc}})$ is understood to be evaluated for the same $(\theta$, $E_{\text{inc}})$. If any noise present in the measurement is a constant fraction of the signal over the whole electron energy spectrum, i.e. $Z/T = Z_{\epsilon_0}/T_{\epsilon_0}$, then, even for cases where $Z \ne 0$, $\tilde{R}^\text{meas} = R$ and the expression for the corrected, normalized contrast reduces to the true contrast, 
\begin{equation}
    \tilde{C}^\text{meas} = C
\end{equation}

An example phase sweep in unprocessed form and the correction of both slow drift and higher frequency fluctuations with the corrected peak normalization scheme is shown in Fig. 3. As Fig. 3 clearly shows, the corrected normalization scheme is effective in removing electron count fluctuations due to drift while preserving the general shape of the curve. 

The maximum energy modulation for a given $\theta$ is robust to beam misalignment. Because the maximum energy modulation occurs at the edge of the lens channel, where the evanescent field is strongest, the maximum energy modulation for a given $\theta$ does not change with small misalignments of the electron beam - so long as the beam remains partially within the channel, some electrons from the beam experience the largest possible energy modulation provided by the optical mode. In this regime, the maximum energy modulation depends solely on the amplitude and phase of the drive lasers, whose fluctuations are small compared to the electron count fluctuations introduced by beam pointing instability. 

Shown below each panel in Fig. 3 is the electron energy spectrum as a function of frame number (time) as the laser phase $\theta$ is linearly increased. The correlation between the contrast and the energy broadening is supported by analytical considerations (See Eq. 4 in the main paper), and is clearly seen in Fig. 3. This provides a good sanity check for our normalization scheme. 

Restating the key result, if the noise is evenly distributed in electron energy space, then the corrected, normalized contrast $\tilde{C}$ is equal to the true contrast $C$, and the correct (de-noised) value of contrast for each data point can be recovered through the peak normalization scheme without use of a distinct reference beam.

\end{document}